\documentclass[aps,prb,twocolumn,groupedaddress]{revtex4-1}
\usepackage{graphics}
\usepackage{graphicx}
\usepackage{dcolumn}
\usepackage{bm}
\usepackage{float}
\usepackage{color,hyperref}

\begin{document}

\title{Thermoelectric Properties of Polycrystalline NiSi$_3$P$_4$}

\author{Andrew F. May}
\email{mayaf@ornl.gov}
\author{Michael A. McGuire}
\author{Hsin Wang}
\affiliation{Materials Science and Technology Division, Oak Ridge National Laboratory, Oak Ridge, TN 37831}

\begin{abstract}
The Hall and Seebeck coefficients, electrical resistivity and thermal conductivity of polycrystalline NiSi$_3$P$_4$ were characterized from 2 to 775\,K.  Undoped NiSi$_3$P$_4$ behaves like a narrow gap semiconductor, with activated electrical resistivity $\rho$ below room temperature and a large Seebeck coefficient of $\sim$ 400\,$\mu$V/K at 300\,K.  Attempts to substitute boron for silicon resulted in the production of extrinsic holes, yielding moderately-doped semiconductor behavior with $\rho$ increasing with increasing temperature above $\sim$ 150\,K.  Hall carrier densities are limited to approximately 5$\times$10$^{19}$cm$^{-3}$ at 200\,K, which would suggest the solubility limit of boron is reached if boron is indeed incorporated into the lattice.  These extrinsic samples have a Hall mobility of $\sim$12\,cm$^2$/V/s at 300\,K, and a parabolic band equivalent effective mass of $\sim$3.5 times the free electron mass.  At 700\,K, the thermoelectric figure of merit $zT$ reaches $\sim$0.1. Further improvements in thermoelectric performance would require reaching higher carrier densities, as well as a mechanism to further reduce the lattice thermal conductivity, which is $\sim$ 5\,W/m/K at 700\,K.  Alloying in Ge results in a slight reduction of the thermal conductivity at low temperatures, with little influence observed at higher temperatures.
\end{abstract}

%\thanks{Copyright (2013) American Institute of Physics. This article may be downloaded for personal use only. Any other use requires prior permission of the author and the American Institute of Physics.  The following article appeared in (J. Appl. Phys. 113, 103707 (2013)) and may be found at (\href{http://dx.doi.org/10.1063/1.4794992}{this link}).}

\maketitle

\section*{}\small \onecolumngrid
Copyright (2013) American Institute of Physics. This article may be downloaded for personal use only. Any other use requires prior permission of the author and the American Institute of Physics.  The following article appeared in (J. Appl. Phys. 113, 103707 (2013)) and may be found at (\href{http://dx.doi.org/10.1063/1.4794992}{this link})
\normalsize \twocolumngrid

\section{Introduction}
Identifying thermoelectric materials composed of environmentally friendly and/or plentiful elements is one current thrust of the thermoelectric community.  Recent successes in this area include several Zintl phases, such as Ca$_3$AlSb$_3$ and Ca$_5$Al$_2$Sb$_6$.\cite{Ca3AlSb3,526Tober}  Simple binaries such as FeSb$_2$ and FeSi may be promising for low-temperature applications.\cite{Bentien_1994,Takahashi_Terasaki_2011,Sales_FeSi_2011}  In addition to environmental concerns and abundance issues, low-density materials are also of interest for applications due to their potential for large specific powers, and the associated reduced mass requirements.  For instance, Mg$_2$Si is an example of a material that possesses all of the above desired properties and good thermoelectric performance.\cite{Mg2SiPRL2012,ZaitsevMg2Si2006,Mg2Si_Bux}  The higher manganese silicides (MnSi$_{2-x}$, $x$$\sim$0.25) are another group of low-cost, abundant materials with promising thermoelectric performance.\cite{Yamada2009,Gelbstein2012}  It is this combination of abundant and low-density materials that brought us to examine another silicide, NiSi$_3$P$_4$, of which little is known beyond the crystallographic structure.  

NiSi$_3$P$_4$ is isostructural with Cu$_3$SbSe$_4$,\cite{NiSi3P4Structure,Cu3SbSe4Structure} which has recently been shown to possess moderate thermoelectric performance.  At the materials level, thermoelectric performance is quantified via the dimensionless figure of merit $zT$=$\alpha^2T$/$\rho\kappa$, where $\alpha$ is the Seebeck coefficient, $\rho$ the electrical resistivity, and $\kappa$ the thermal conductivity.  In Cu$_3$SbSe$_4$, $zT$ reaches between 0.7 and 0.9 near $\sim$650\,K, with the exact value depending on the particular alloying and doping employed.\cite{Cu3SbSe4Yang2011,Cu3SbSe4Skoug2011,Cu3SbSe4Skoug2011B,Cu3SbSe4Skoug2011C}

NiSi$_3$P$_4$ and Cu$_3$SbSe$_4$ are isostructural with the mineral famatinite Cu$_3$SbS$_4$, which is a member of the stannite (Cu$_2$FeSnS$_4$) group.  These materials possess a tetragonal unit cell (space group $I$$\bar{4}$2$m$, No. 121) based on the zinc-blende structure with all atoms being tetrahedrally coordinated (though the tetrahedral coordinations are slightly distorted with bond angles deviating from the ideal 109.5$^{\circ}$).  In Cu$_3$SbSe$_4$, Cu and Sb are tetrahedrally coordinated by Se, whereas in NiSi$_3$P$_4$, Ni and Si are surrounded by P.

In this work, we characterize NiSi$_3$P$_4$, and find it to be a narrow gap semiconductor where holes naturally dominate conduction.  This material is too resistive to yield large thermoelectric performance, and thus doping studies were undertaken.  At first, we attempted to substitute Co for Ni, though this did not appear to produce extrinsic holes.  Thus, in analogy with Si and SiGe alloys, we attempted to use boron as a hole dopant, and nominal compositions NiSi$_{3-x}$B$_x$P$_4$ were found to produce extrinsic hole concentrations.  If boron substitutes for silicon, it does so with a relatively limited solubility, as the maximum hole concentration observed in these samples is approximately 5$\times$10$^{19}$cm$^{-3}$ at 200\,K.  It is also possible the extrinsic hole concentration is associated with off-stoichiometry of the NiSi$_3$P$_4$ phase induced by the nominal compositions employed.  With some modest success at doping, it was clear that the lattice thermal conductivity severely limits the thermoelectric performance.  Thus, we attempted to alloy in 20\% Ge on the Si site, and this resulted in only a small decrease in $\kappa$ (primarily at low temperatures).  The net result is a rather low figure of merit $zT$$\sim$0.1 at 700\,K.  Further improvements in performance may be achieved with higher doping levels, as the material has a relatively high band mass, though a reduction in the lattice thermal conductivity is also necessary.  Therefore, this material is unlikely to be pursued in further detail unless theoretical calculations provide insights into doping or suggest anomalous performance for other regions of the electronic structure.

\section{Methods}
Polycrystalline  samples were prepared by first arc-melting high-purity Ni, Si, B, and Ge as appropriate for the nominal compositions listed in Table \ref{tab1}.  The resulting ingot was hand ground to a fine powder, and combined with red phosphorus in a silica ampoule that was subsequently sealed under vacuum.  The ampoule was heated to 450\,$^{\circ}$C, held at this temperature for 10\,h, then heated to 900\,$^{\circ}$C and held for 100\,h to allow phosphorus to react through the vapor phase.  After cooling, the ampoule was shaken vigorously and heated to 900\,$^{\circ}$C for another $\sim$50-100\,h to ensure full reaction of the phosphorus.  During the first heating, a rate of 25$^{\circ}$C/h was employed, and 50$^{\circ}$C/h was used for the second heat treatment.  Indications of decomposition and significant volatilization of phosphorus were observed when sintering of a cold-pressed pellet at 1000\,$^{\circ}$C was attempted, and thus $\sim$900\,$^{\circ}$C was used as a high temperature limit.

After heating, the powders were hand ground and consolidated with a hot press.  Hot pressing was performed in a graphite die within a graphite furnace, using pressures near 10,000\,psi and a maximum temperature near 900\,$^{\circ}$C$\pm$50\,$^{\circ}$C.  After exposure to the maximum force/temperature for $\sim$1\,h, the force was removed for a one hour stress-free anneal, followed by cooling to 675\,$^{\circ}$C over one hour, at which temperature the furnace was turned off to cool more rapidly.  This procedure resulted in samples with densities greater than 92\% of the crystallographic density.  The samples are relatively stable in air, with only a very slight tarnish appearing after several months.  Grinding was generally done in a helium glove box, though grinding in air prior to hot pressing did not appear to affect the transport properties provided the reaction was complete.  The acrid smell of phosphorus was experienced upon cutting or grinding the final specimens.

The polycrystalline samples were found, by powder x-ray diffraction, to contain mainly the desired NiSi$_3$P$_4$ phase.  The undoped NiSi$_3$P$_4$ sample contained a small amount ($<$2 wt.\%) of Ni$_2$Si. Rietveld refinement of the room temperature powder diffraction data using the published structural model\cite{NiSi3P4Structure} yielded lattice parameters of $a$=5.155(1)$\AA$ and $c$=10.345(1)$\AA$.  Boron substitution resulted in a negligible change in lattice parameters.  With increasing nominal concentrations of boron, the impurity phase(s) changed from Ni$_2$Si to BP and NiSi, with up to 3\% and 6.4\%, respectively, observed in sample 4,B in Table \ref{tab1}.  As discussed below, transport measurements suggest that the boron solubility limit was reached by the nominal composition of NiSi$_{2.98}$B$_{0.02}$P$_4$.  It is also possible that the phase width of NiSi$_3$P$_4$ is reached at this nominal composition and boron does not incorporate into the lattice. In the sample with nominal composition NiSi$_2$Ge$_{0.6}$B$_{0.4}$P$_4$, the diffraction revealed about 87\% purity of the main phase, with BP (7\,wt.\%) and Ge (6 \,wt.\%) as the primary secondary phases.  The lattice parameters indicate some Ge incorporated into the lattice, though, with refinement yielding $a$=5.169(1)$\AA$ and $c$=10.367(1)$\AA$.

\begin{table}%
\caption{Nominal compositions and lattice parameters of the NiSi$_3$P$_4$ and doped-NiSi$_3$P$_4$ samples presented in this study.}
\begin{tabular}{rlcc}
\hline
sample ID   & nominal composition                  & $a$       & $c$ \\ 
            &                                      & $\AA$     & $\AA$ \\ 
\hline
1           & NiSi$_3$P$_4$                        & 5.155(1)  &  10.345(1) \\  %205D
2, B        & NiSi$_{2.98}$B$_{0.02}$P$_4$         & 5.155(1)  &  10.345(1) \\  %211B 
3, B        & NiSi$_{2.90}$B$_{0.10}$P$_4$         & 5.157(1)  &  10.347(1) \\  %215B
4, B        & NiSi$_{2.80}$B$_{0.20}$P$_4$         & 5.154(1)  &  10.342(1) \\  %215C
5, B \& Ge  & NiSi$_{2}$Ge$_{0.6}$B$_{0.40}$P$_4$  & 5.169(1)  &  10.367(1) \\  %227B
\hline
\end{tabular}
\label{tab1}
\end{table}

Transport data suggested very high hole concentrations are necessary to achieve large thermoelectric performance, and thus we attempted to synthesize samples with large concentrations of boron.   The secondary phases seem to have little influence on the transport properties, as similar results were obtained on the different samples.  To allow for an appreciation of the variations in properties across batches, we report data on several boron containing samples that possess similar transport properties and thus likely possess similar amounts of boron in the primary phase.

\begin{figure}[!ht]
	\centering
\includegraphics[width=3in]{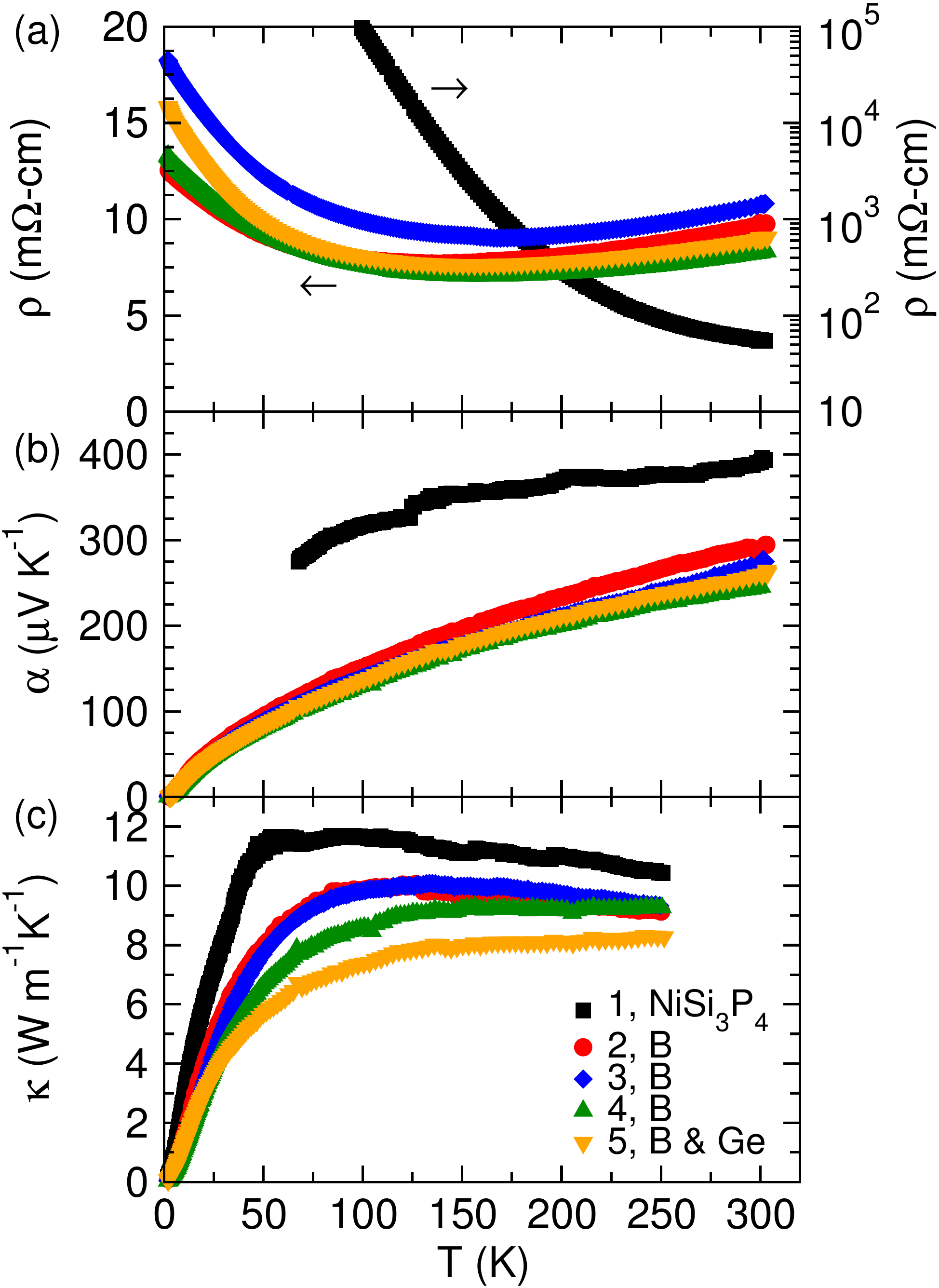}
\caption{(color online) (a) Electrical resistivity, (b) Seebeck coefficient, and (c) thermal conductivity of polycrystalline NiSi$_3$P$_4$, as well as B- and Ge-doped compositions as listed in Table \ref{tab1}.}
	\label{fig:TTO}
\end{figure}

Seebeck coefficient $\alpha$, electrical resistivity $\rho$, and Hall coefficient $R_{H}$ measurements were utilized to investigate the electrical properties; the Hall carrier density is $n_H$=1/$R_He$.  Below 300\,K, these measurements were performed in a Quantum Design Physical Property Measurement System (PPMS).  The thermoelectric measurements, including the thermal conductivity $\kappa$, were performed using the PPMS Thermal Transport Option (TTO), and the Hall effect was measured using the resistivity option on the PPMS.  For TTO measurements, gold-plated copper leads were attached to the sample using silver epoxy (H20E Epo-Tek), and silver paste (DuPont 4929N) was utilized to obtain low resistance contacts for Hall effect measurements performed using 0.025\,mm Pt wires. Hall coefficients were obtained from a fit of the Hall resistance versus magnetic field, with maximum fields of $\pm$6\,Tesla employed.  

High temperature thermoelectric performance was assessed using an ULVAC ZEM-3 M8 to measure $\alpha$ and $\rho$. The thermal conductivity $\kappa$ (above room temperature) was calculated via $\kappa = D_T \times C_P \times d$ where $d$ is the density. The specific heat capacity $C_p$ was approximated using the Dulong Petit limit of $C_v$= 3R, calculated according to nominal composition.  The differences between $C_v$ and $C_P$ are assumed to be negligible at the temperatures of interest, and this estimation likely results in a slight overestimation of $zT$ at the highest temperatures.  The thermal diffusivity $D_T$ was obtained on $\sim$10\,mm diameter discs between 1.5 and 2\,mm thick, in an Anter FL-5000 with a graphite furnace; data analysis followed ASTM 1461 for flash diffusivity.  No sign of oxidation occurred during the high temperature measurements, and data collected on cooling were consistent with those collected on heating.

\section{Results and Discussion}

NiSi$_3$P$_4$ was found to behave similar to a narrow gap semiconductor with a low concentration of charge carriers.  As shown in Figure \ref{fig:TTO}, activated electrical conduction is observed for NiSi$_3$P$_4$ below room temperature.  Between 150 and 250\,K, a simple activation energy was obtained from fitting to $\rho$=$\rho_0$Exp[$E_a$/kT], resulting in $E_a$=0.11\,eV, which is equivalent to a fundamental band gap of 0.22\,eV.  However, the fit to this simple activated behavior becomes poor when wider temperature ranges are considered.

\begin{figure}
	\centering
\includegraphics[width=3in]{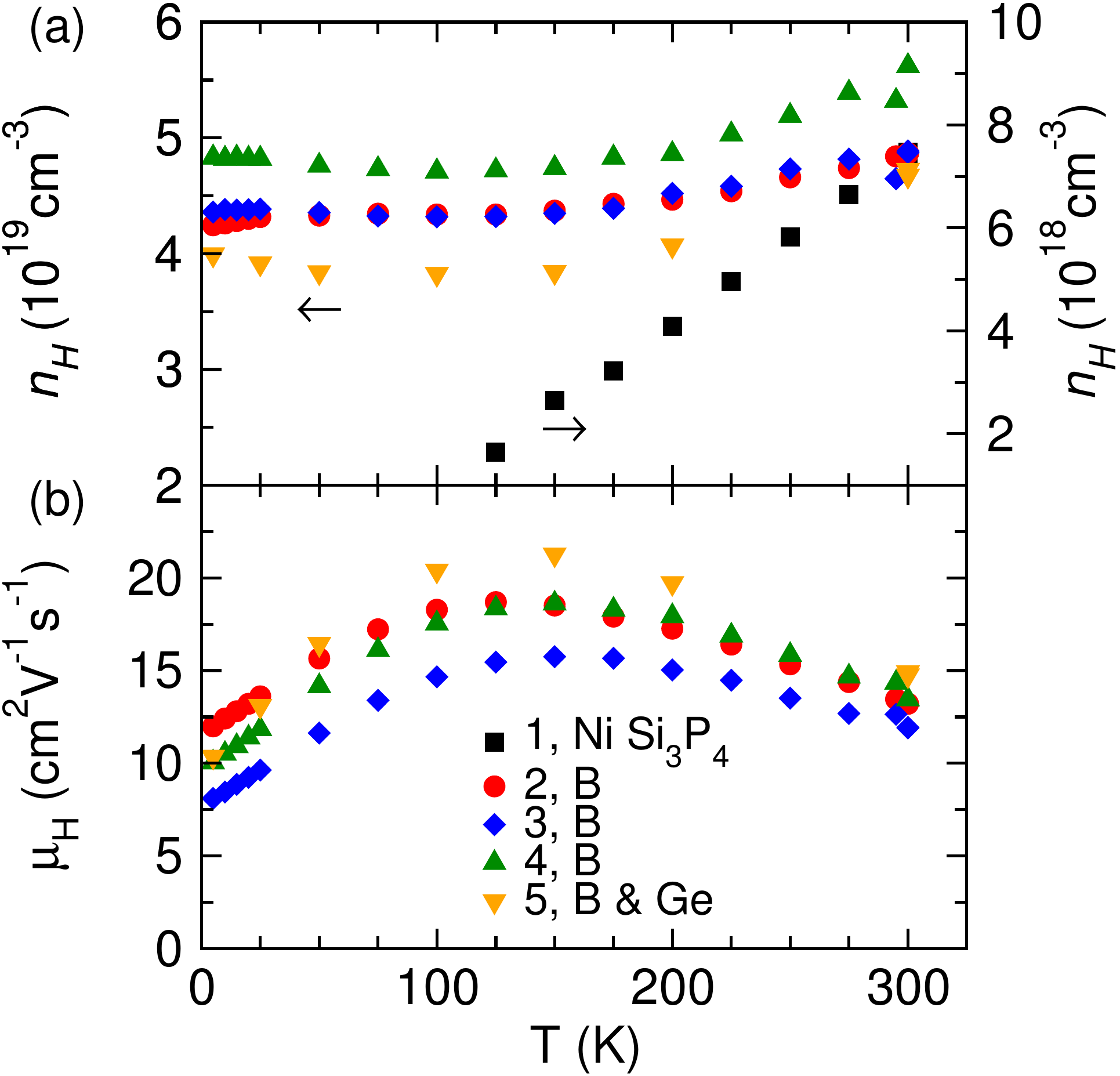}
\caption{(color online) (a) Hall carrier concentration and (b) Hall mobility of polycrystalline NiSi$_3$P$_4$, as well as B- and Ge-doped compositions as listed in Table \ref{tab1}.}
	\label{fig:Hall}
\end{figure}

Consistent with the semiconducting-like behavior, a large Seebeck coefficient is observed in NiSi$_3$P$_4$, as shown in Figure \ref{fig:TTO}b.  Also consistent with this behavior, the Hall carrier density $n_H$ increases with increasing temperature for NiSi$_3$P$_4$, as shown in Figure \ref{fig:Hall}a.  We emphasize that this is a single band $n_H$, and may not reflect the true carrier density if electrons are also contributing to electrical transport.

In order to probe the potential for thermoelectric performance, we attempted to hole-dope NiSi$_3$P$_4$.  We first tried the substitution of Co for Ni, but this sample had a higher resistivity than undoped NiSi$_3$P$_4$ at 300\,K, and also displayed activated conduction below room temperature.  We then found that boron successfully increased the hole concentration in samples with nominal compositions of NiSi$_{3-x}$B$_x$P$_4$.  However, synthesis with $x$=0.02, 0.10, 0.20, 0.40 all yielded similar transport properties, as observed in Figure \ref{fig:TTO}.  One explanation for this trend is that the solubility limit of boron has been reached. 

Hole-doped samples of NiSi$_{3-x}$B$_x$P$_4$ behave metallic above $\sim$ 150\,K with $\rho$ increasing with increasing $T$.  Similarly, the Seebeck coefficients of these samples increase with increasing $T$, consistent with extrinsically doped samples.  The Seebeck coefficients remain rather large in these doped samples, with $\alpha$$\sim$250$\mu$V/K obtained at 300\,K.

The Hall carrier concentration and mobility $\mu_H$ of boron- and boron/germanium-doped NiSi$_{3}$P$_4$ are shown in Figure \ref{fig:Hall}.  As observed, the values of $n_H$ are relatively temperature independent, and fairly similar for the different nominal boron concentrations. This is consistent with similar $\rho$ and $\alpha$ in the various samples.  This result suggests that, if boron substitutes for silicon, the solubility limit of boron has been reached.  If lattice-incorporated boron atoms are fully ionized and contribute one hole to the system, then a solubility limit of $\sim$0.2\%B per silicon is obtained from the Hall carrier concentration of approximately 5$\times$10$^{19}$holes/cm$^{3}$ at 200\,K.  This seems reasonable considering the solubility limits of boron in silicon, which are $\sim$ 2$\times$10$^{19}$B/cm$^{3}$ at 700$^{\circ}$C and  $\sim$ 1$\times$10$^{20}$B/cm$^{3}$ at 900$^{\circ}$C.\cite{BoronSilicon}  The small increase of $n_H$ above $\sim$200\,K may be the result of some carriers being activated from other impurities or across the band gap.

The hole mobility of boron-doped NiSi$_3$P$_4$ is found to be approximately 12\,cm$^2$/V/s at 300\,K.  The mobility increases with decreasing temperature, consistent with a carrier mean free path limited by phonon scattering.  However, this behavior only occurs down to $\sim$125\,K, where the mobility reaches a maximum of approximately 18\,cm$^2$/V/s.  While $\mu_H$ decreases upon cooling below 125\,K, $n_H$ remains relatively constant, which may suggest that additional carrier scattering mechanisms are important at low $T$.  Ionized impurity scattering generally leads to a mobility that increases as $T^{3/2}$, though the low-T rise in $\mu_H$ occurs more slowly than $T^{3/2}$ in  boron-doped NiSi$_3$P$_4$.  It is also possible that Anderson localization occurs due to disorder in the lattice, causing a mobility gap that leads to an increase in $\rho$ at low-$T$.  Also, we note that the high-temperature decay occurs more slowly than the $T^{-3/2}$ behavior expected for a lightly doped semiconductor (or even the $T^{-1}$ behavior expected for a degenerate semiconductor).  Therefore, it is possible impurity band conduction is important in these materials or grain boundary scattering influences transport.

The hole mobility of boron-doped NiSi$_3$P$_4$ is less than that of isostructural Cu$_3$SbSe$_4$.  The mobility of undoped Cu$_3$SbSe$_4$ is $\sim$41\,cm$^2$/V/s at 300\,K, with a hole concentration of 7.5$\times$10$^{18}$cm$^{-3}$.\cite{Cu3SbSe4Yang2011}  The mobility of undoped NiSi$_3$P$_4$ is not shown in Figure \ref{fig:Hall} because the single band model likely fails for the undoped composition.  At a more comparable hole concentration of 6.3$\times$10$^{19}$cm$^{-3}$ in Cu$_3$Sb$_{0.975}$Sn$_{0.025}$Se$_4$ the mobility is approximately 35\,cm$^2$/V/s, and it is 31\,cm$^2$/V/s in Cu$_3$Sb$_{0.95}$Sn$_{0.05}$Se$_4$ at 300\,K.\cite{Cu3SbSe4Yang2011}  Given that the carrier mobility in doped NiSi$_3$P$_4$ is dominated by phonon scattering at room temperature, it seems unlikely that the boron dopant is leading to this large difference in carrier mobility.  Therefore, the hole mobility in NiSi$_3$P$_4$ appears to be inherently lower than in the copper-based analogues.

To better understand the differences between transport in NiSi$_3$P$_4$ and Cu$_3$SbSe$_4$, we consider a parabolic band analysis of the Hall and Seebeck coefficients (Equations \ref{eqn:alpha}--\ref{eqn:F}).  Analysis of the Hall and Seebeck coefficient data allows the carrier mass $m^*$ to be obtained.  When phonon scattering limits the relaxation time of charge carriers, the mobility is expected to scale as $\mu$$\propto$$m^{*-5/2}$.  Therefore, even small changes in the band mass can have significant influence on carrier transport.

Within a single parabolic band model, the Seebeck coefficient is a function of the reduced electrochemical potential $\eta$ and scattering parameter $\lambda$:

\begin{center}
\begin{equation}
\alpha=\frac{k}{e}\left(\frac{(2+\lambda)F_{\lambda+1}(\eta)}{(1+\lambda)F_{\lambda}(\eta)}-\eta\right).
\label{eqn:alpha}
\end{equation}
\end{center}

\noindent The scattering parameter $\lambda$ is used to incorporate the energy dependence of the carrier relaxation time, $\tau=\tau_0\epsilon^{\lambda-0.5}$.  The value of $\lambda$ influences $\alpha$, but $\tau_0$ does not.  In this analysis, we assume acoustic phonon scattering limits $\tau$, corresponding to $\lambda$=0.

The Hall carrier density, $n_H$=1/$R_H$e where $R_H$ is the Hall coefficient, is given by 

\begin{center}
\begin{equation}
n_{H}=4\pi\left(\frac{2 m^*kT}{h^{2}}\right)^{3/2}\frac{F_{1/2}(\eta)}{r_{H}}.
\label{eqn:n}
\end{equation}
\end{center}

\noindent Analysis of $n_H$ allows the effective mass $m^*$ to be determined for a given $\eta$ obtained from analysis of the Seebeck coefficient.  Carrier scattering influences $n_H$ through the Hall factor $r_{H}$, which is a function of $\eta$ and $\lambda$ similar to the Seebeck coefficient:

\begin{center}
\begin{equation}
r_{H}=\frac{3}{2}F_{1/2}(\eta)\frac{(1/2+2\lambda)F_{2\lambda-1/2}(\eta)}{(1+\lambda)^2F_{\lambda}^{2}(\eta)}.
\label{eqn:rH}
\end{equation}
\end{center}

\noindent For simplification purposes, these expressions utilize the Fermi integrals $F_{j}(\eta$) defined by
\begin{center}
\begin{equation}
F_{j}(\eta)=\int_0^{\infty} \frac{\xi^{j}\,d\xi}{1+Exp[\xi-\eta]}.
\label{eqn:F}
\end{equation}
\end{center}

\begin{figure}
	\centering
\includegraphics[width=3in]{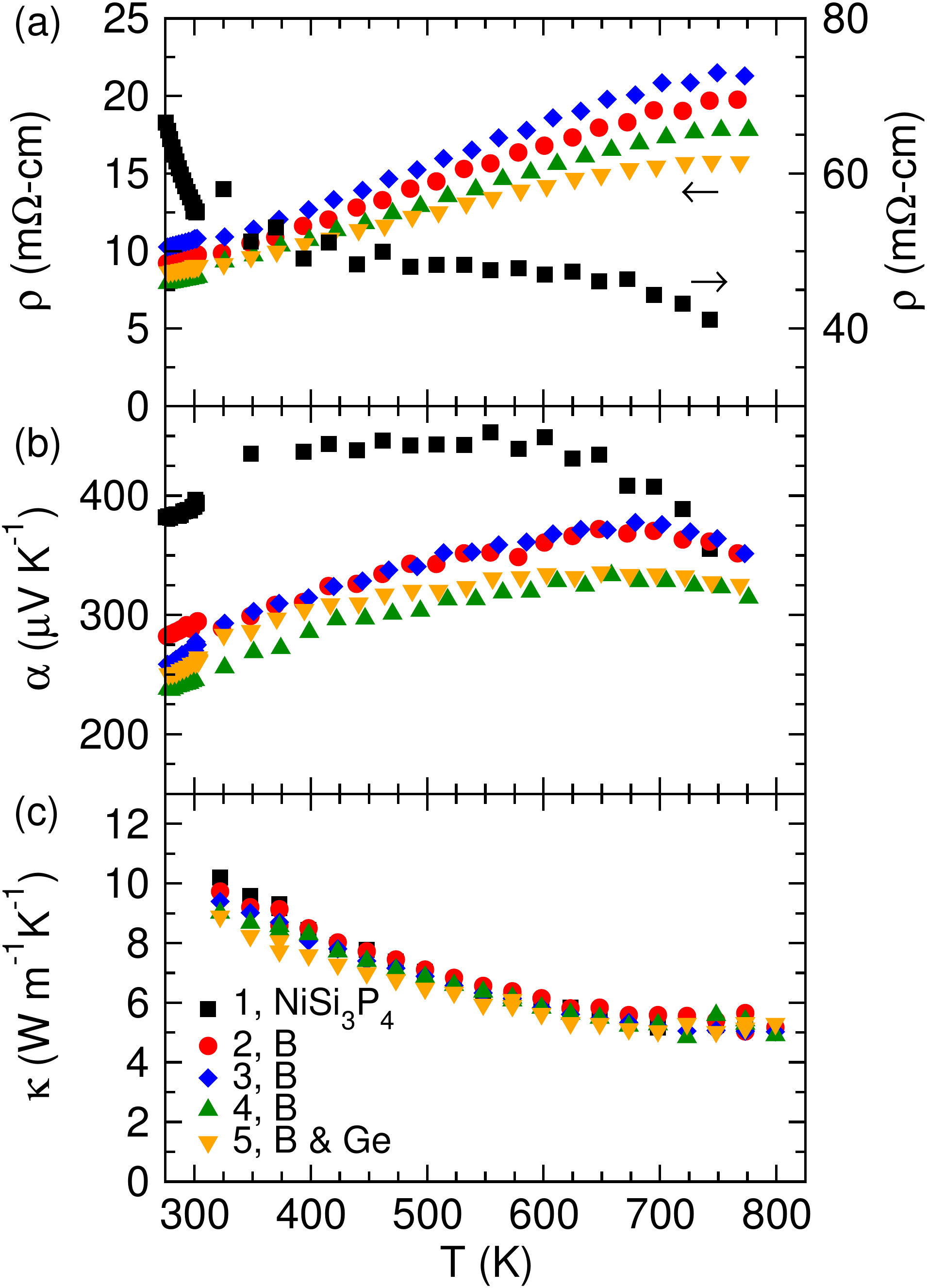}
\caption{(color online) High temperature (a) electrical resistivity, (b) Seebeck coefficient, and (c) thermal conductivity of polycrystalline NiSi$_3$P$_4$, as well as B- and Ge-doped compositions as listed in Table \ref{tab1}.}
	\label{fig:ZEM}
\end{figure}

Employing the above equations, we find that the effective mass is approximately 3.5$\pm$0.5$m_e$ in boron-doped NiSi$_3$P$_4$ at room temperature; $m_e$ is the free electron mass. At 200\,K, where $n_H$ is still relatively independent of temperature, this process yields 2.9-3.7$m_e$ for the various samples of boron-doped NiSi$_3$P$_4$.  From literature values for $n_H$ and $\alpha$ in Cu$_3$SbSe$_4$ (corresponding to the $\mu$ values noted above),\cite{Cu3SbSe4Yang2011} we obtain $0.45m_e<m*<1.3m_e$ depending on the particular sample.  Therefore, it is clear that the band mass is much larger in NiSi$_3$P$_4$ than in Cu$_3$SbSe$_4$, and this likely contributes to the lower mobility.  This large band mass results in a relatively large Seebeck coefficient in NiSi$_3$P$_4$, but the impact on carrier mobility is detrimental to the thermoelectric performance of NiSi$_3$P$_4$.

The thermal conductivity of NiSi$_3$P$_4$ is approximately 10\,W/m/K at 300\,K.  Boron doping reduces the low-temperature value of $\kappa$ slightly.  In an attempt to further reduce $\kappa$, we substituted Ge for Si, in addition to the boron doping.  This results in some additional suppression of $\kappa$ at low-$T$, though little change is observed at higher temperatures.  Interestingly, $\kappa$ does not display much temperature dependence below room temperature, even in undoped NiSi$_3$P$_4$.  This suggests there may be some disorder in the system, or perhaps grain boundaries or secondary phases are reducing $\kappa$ at low-$T$.  We note that due to the large values of $\rho$, the lattice thermal conductivity $\kappa_L$ dominates $\kappa$ for all samples below $\sim$600\,K.

The high temperature transport properties are plotted in Figure \ref{fig:ZEM}.  Interestingly, the decrease in $\rho$ with increasing $T$ for undoped NiSi$_3$P$_4$ turns into a nearly temperature independent $\rho$ above $\sim$400\,K, and the corresponding Seebeck coefficient is large and relatively temperature independent.  The boron-doped samples display typical doped semiconductor behavior at high temperatures, with the resistivity and Seebeck coefficients increasing with increasing temperature until the effects of minority carrier activation are observed near 650\,K.  The thermal conductivity decreases with increasing $T$ in all samples, consistent with the dominant role of phonon-phonon scattering in these crystalline materials.  As noted above, the substitution of 20\% Ge (nominal composition) did not result in a significant reduction in $\kappa$ above 300\,K.  For comparison, the thermal conductivity of Cu$_3$SbS$_4$ reaches approximately 1.5\,W/m/K at 600\,K.\cite{Cu3SbSe4Yang2011}

\begin{figure}
	\centering
\includegraphics[width=3in]{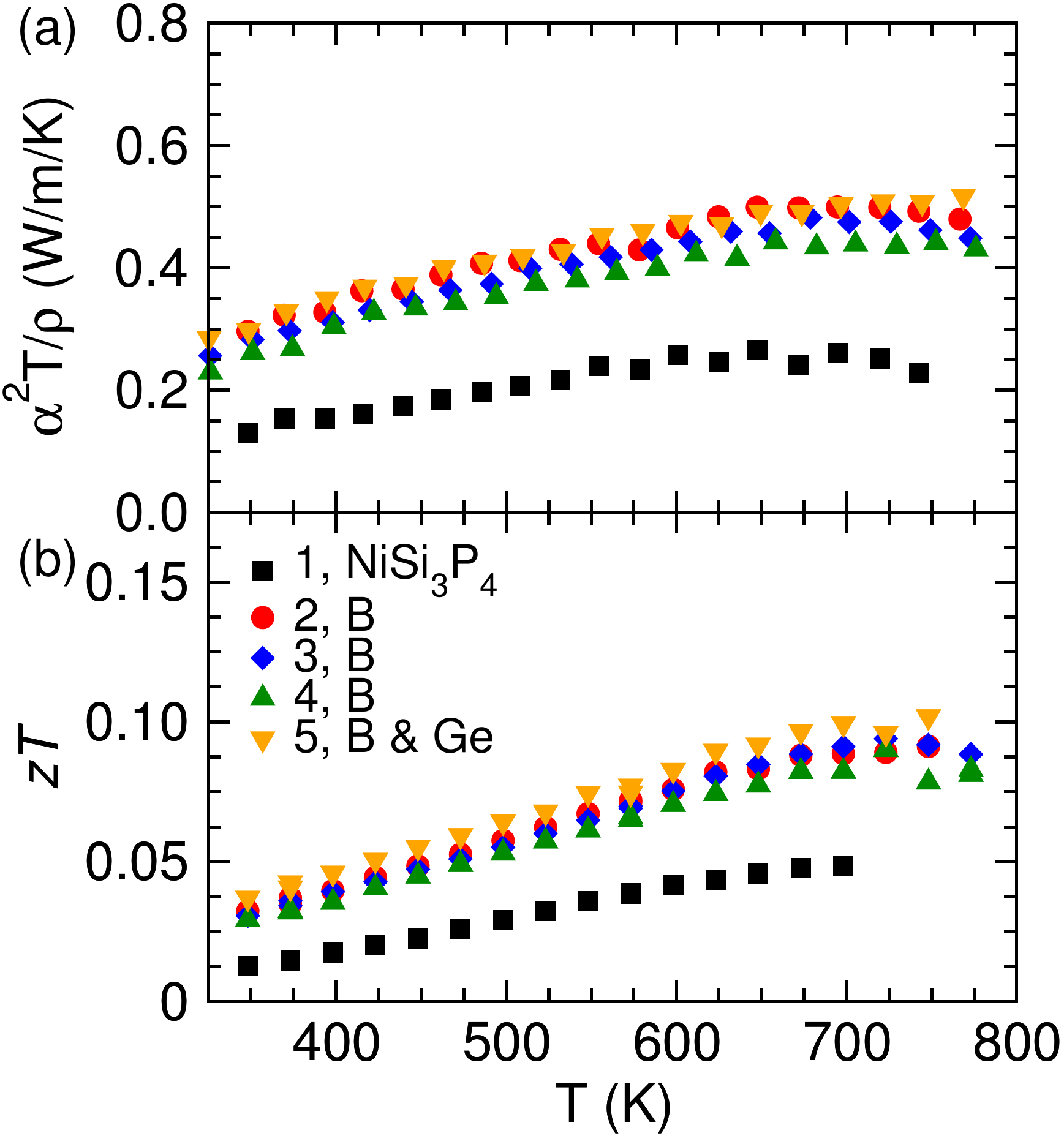}
\caption{(color online) (a) The thermoelectric power factor times temperature and (b) dimensionless figure of merit for polycrystalline NiSi$_3$P$_4$, as well as B- and Ge-doped compositions as listed in Table \ref{tab1}.}
	\label{fig:zT}
\end{figure}

The maximum in $\alpha(T)$, which is relatively unpronounced in these materials, can be utilized to estimate the thermal band gap by $E_g$=2$\alpha_{max}T_{max}$.\cite{EgEstimate}  This analysis suggests that the band gap is between 0.43 and 0.50\,eV, which is about two times larger than that obtained from the analysis of $\rho(T)$ at lower temperatures.  It is possible that one of these estimates of $E_g$ is influenced by the activation of carriers from an impurity band.  Without high temperature Hall effect data the source for this discrepancy is difficult to isolate.

The thermoelectric power factor, $\alpha^2T/\rho$ is plotted in Figure \ref{fig:zT} along with the figure of merit $zT$.  Relatively low thermoelectric performance is observed in these materials.  This is due in part to the relatively high thermal conductivity for the corresponding mobility values.  Boron-doping greatly improves both the power factor and $zT$, and additional improvements would likely come with higher doping levels.  This is particularly the case for $p$-type NiSi$_3$P$_4$, because it is a large band mass material and the optimum carrier density increases with increasing band mass and increasing lattice thermal conductivity.\cite{Ioffe,ComplexTE}  As noted, however, boron substitution only permits modest carrier densities to be reached, and using a transition metal dopant was not successful in increasing the hole concentration.  It is possible that altering the Si/P ratio would permit higher doping levels to be achieved, and may also allow $n$-type behavior to be probed.  However, it seems unlikely that this material will achieve the $zT$ values observed in isostructural copper compounds, which have higher mobility due to lower band masses, as well as lower lattice thermal conductivities.

\subsection{Summary}

Polycrystalline NiSi$_3$P$_4$ was found to be a narrow gap semiconductor, with an activation energy for the electrical resistivity on the order of 100\,meV below room temperature.  Boron-doping is successful in producing extrinsic holes, though the corresponding hole concentration is limited to approximately 5$\times$10$^{19}$cm$^{-3}$, suggesting a possible solubility limit of boron in NiSi$_3$P$_4$.  The transport properties of the boron-doped samples behave as expected for moderately doped semiconductors, and parabolic band analysis suggests a fairly high band mass of $\sim$3$m_e$.  The thermoelectric figure of merit only reaches $\sim$0.1 at 700\,K, due in large part to a lower than desired hole mobility and carrier concentration, and a moderately large thermal conductivity.  Further improvements in thermoelectric performance may be possible with a different doping scheme or through additional phonon scattering, but are unlikely to yield large thermoelectric efficiencies. 

\section{Acknowledgements}

This research was supported by the U. S. Department of Energy, Office of Basic Energy Sciences, Materials Sciences and Engineering Division (A.F.M., M.A.M), and the US Department of Energy, EERE, Vehicle Technologies, Propulsion Materials Program (H.W.).


\begin{thebibliography}{20}
\expandafter\ifx\csname natexlab\endcsname\relax\def\natexlab#1{#1}\fi
\expandafter\ifx\csname bibnamefont\endcsname\relax
  \def\bibnamefont#1{#1}\fi
\expandafter\ifx\csname bibfnamefont\endcsname\relax
  \def\bibfnamefont#1{#1}\fi
\expandafter\ifx\csname citenamefont\endcsname\relax
  \def\citenamefont#1{#1}\fi
\expandafter\ifx\csname url\endcsname\relax
  \def\url#1{\texttt{#1}}\fi
\expandafter\ifx\csname urlprefix\endcsname\relax\def\urlprefix{URL }\fi
\providecommand{\bibinfo}[2]{#2}
\providecommand{\eprint}[2][]{\url{#2}}

\bibitem[{\citenamefont{Zevalkink et~al.}(2011)\citenamefont{Zevalkink,
  Toberer, Zeier, Flage-Larsen, and Snyder}}]{Ca3AlSb3}
\bibinfo{author}{\bibfnamefont{A.}~\bibnamefont{Zevalkink}},
  \bibinfo{author}{\bibfnamefont{E.~S.} \bibnamefont{Toberer}},
  \bibinfo{author}{\bibfnamefont{W.~G.} \bibnamefont{Zeier}},
  \bibinfo{author}{\bibfnamefont{E.}~\bibnamefont{Flage-Larsen}},
  \bibnamefont{and} \bibinfo{author}{\bibfnamefont{G.~J.}
  \bibnamefont{Snyder}}, \bibinfo{journal}{Energy Environ. Sci.}
  \textbf{\bibinfo{volume}{4}}, \bibinfo{pages}{510} (\bibinfo{year}{2011}).

\bibitem[{\citenamefont{Toberer et~al.}(2010)\citenamefont{Toberer, Zevalkink,
  Crisosto, and Snyder}}]{526Tober}
\bibinfo{author}{\bibfnamefont{E.~S.} \bibnamefont{Toberer}},
  \bibinfo{author}{\bibfnamefont{A.}~\bibnamefont{Zevalkink}},
  \bibinfo{author}{\bibfnamefont{N.}~\bibnamefont{Crisosto}}, \bibnamefont{and}
  \bibinfo{author}{\bibfnamefont{G.~J.} \bibnamefont{Snyder}},
  \bibinfo{journal}{Adv. Funct. Mater.} \textbf{\bibinfo{volume}{20}},
  \bibinfo{pages}{4375} (\bibinfo{year}{2010}).

\bibitem[{\citenamefont{Bentien et~al.}(2007)\citenamefont{Bentien, Johnsen,
  Madsen, Iversen, and Steglich}}]{Bentien_1994}
\bibinfo{author}{\bibfnamefont{A.}~\bibnamefont{Bentien}},
  \bibinfo{author}{\bibfnamefont{S.}~\bibnamefont{Johnsen}},
  \bibinfo{author}{\bibfnamefont{G.~K.~H.} \bibnamefont{Madsen}},
  \bibinfo{author}{\bibfnamefont{B.~B.} \bibnamefont{Iversen}},
  \bibnamefont{and} \bibinfo{author}{\bibfnamefont{F.}~\bibnamefont{Steglich}},
  \bibinfo{journal}{E P L} \textbf{\bibinfo{volume}{80}},
  \bibinfo{pages}{17008} (\bibinfo{year}{2007}).

\bibitem[{\citenamefont{Takahashi et~al.}(2011)\citenamefont{Takahashi,
  Okazaki, Yasui, and Terasaki}}]{Takahashi_Terasaki_2011}
\bibinfo{author}{\bibfnamefont{H.}~\bibnamefont{Takahashi}},
  \bibinfo{author}{\bibfnamefont{R.}~\bibnamefont{Okazaki}},
  \bibinfo{author}{\bibfnamefont{Y.}~\bibnamefont{Yasui}}, \bibnamefont{and}
  \bibinfo{author}{\bibfnamefont{I.}~\bibnamefont{Terasaki}},
  \bibinfo{journal}{Phys. Rev. B} \textbf{\bibinfo{volume}{84}},
  \bibinfo{pages}{205215} (\bibinfo{year}{2011}),
  \urlprefix\url{http://link.aps.org/doi/10.1103/PhysRevB.84.205215}.

\bibitem[{\citenamefont{Sales et~al.}(2011)\citenamefont{Sales, Delaire,
  McGuire, and May}}]{Sales_FeSi_2011}
\bibinfo{author}{\bibfnamefont{B.~C.} \bibnamefont{Sales}},
  \bibinfo{author}{\bibfnamefont{O.}~\bibnamefont{Delaire}},
  \bibinfo{author}{\bibfnamefont{M.~A.} \bibnamefont{McGuire}},
  \bibnamefont{and} \bibinfo{author}{\bibfnamefont{A.~F.} \bibnamefont{May}},
  \bibinfo{journal}{Phys. Rev. B} \textbf{\bibinfo{volume}{83}},
  \bibinfo{pages}{125209} (\bibinfo{year}{2011}),
  \urlprefix\url{http://link.aps.org/doi/10.1103/PhysRevB.83.125209}.

\bibitem[{\citenamefont{Liu et~al.}(2012)\citenamefont{Liu, Tan, Yin, Liu,
  Tang, Shi, Zhang, and Uher}}]{Mg2SiPRL2012}
\bibinfo{author}{\bibfnamefont{W.}~\bibnamefont{Liu}},
  \bibinfo{author}{\bibfnamefont{X.}~\bibnamefont{Tan}},
  \bibinfo{author}{\bibfnamefont{K.}~\bibnamefont{Yin}},
  \bibinfo{author}{\bibfnamefont{H.}~\bibnamefont{Liu}},
  \bibinfo{author}{\bibfnamefont{X.}~\bibnamefont{Tang}},
  \bibinfo{author}{\bibfnamefont{J.}~\bibnamefont{Shi}},
  \bibinfo{author}{\bibfnamefont{Q.}~\bibnamefont{Zhang}}, \bibnamefont{and}
  \bibinfo{author}{\bibfnamefont{C.}~\bibnamefont{Uher}},
  \bibinfo{journal}{Phys. Rev. Lett.} \textbf{\bibinfo{volume}{108}},
  \bibinfo{pages}{166601} (\bibinfo{year}{2012}),
  \urlprefix\url{http://link.aps.org/doi/10.1103/PhysRevLett.108.166601}.

\bibitem[{\citenamefont{Zaitsev et~al.}(2006)\citenamefont{Zaitsev, Fedorov,
  Gurieva, Eremin, Konstantinov, Samunin, and Vedernikov}}]{ZaitsevMg2Si2006}
\bibinfo{author}{\bibfnamefont{V.~K.} \bibnamefont{Zaitsev}},
  \bibinfo{author}{\bibfnamefont{M.~I.} \bibnamefont{Fedorov}},
  \bibinfo{author}{\bibfnamefont{E.~A.} \bibnamefont{Gurieva}},
  \bibinfo{author}{\bibfnamefont{I.~S.} \bibnamefont{Eremin}},
  \bibinfo{author}{\bibfnamefont{P.~P.} \bibnamefont{Konstantinov}},
  \bibinfo{author}{\bibfnamefont{A.~Y.} \bibnamefont{Samunin}},
  \bibnamefont{and} \bibinfo{author}{\bibfnamefont{M.~V.}
  \bibnamefont{Vedernikov}}, \bibinfo{journal}{Phys. Rev. B}
  \textbf{\bibinfo{volume}{74}}, \bibinfo{pages}{045207}
  (\bibinfo{year}{2006}),
  \urlprefix\url{http://link.aps.org/doi/10.1103/PhysRevB.74.045207}.

\bibitem[{\citenamefont{Bux et~al.}(2011)\citenamefont{Bux, Yeung, Toberer,
  Snyder, Kaner, and Fleurial}}]{Mg2Si_Bux}
\bibinfo{author}{\bibfnamefont{S.~K.} \bibnamefont{Bux}},
  \bibinfo{author}{\bibfnamefont{M.~T.} \bibnamefont{Yeung}},
  \bibinfo{author}{\bibfnamefont{E.~S.} \bibnamefont{Toberer}},
  \bibinfo{author}{\bibfnamefont{G.~J.} \bibnamefont{Snyder}},
  \bibinfo{author}{\bibfnamefont{R.~B.} \bibnamefont{Kaner}}, \bibnamefont{and}
  \bibinfo{author}{\bibfnamefont{J.-P.} \bibnamefont{Fleurial}},
  \bibinfo{journal}{J. Mater. Chem.} \textbf{\bibinfo{volume}{21}},
  \bibinfo{pages}{12259} (\bibinfo{year}{2011}).

\bibitem[{\citenamefont{Itohi and Yamada}(2009)}]{Yamada2009}
\bibinfo{author}{\bibfnamefont{T.}~\bibnamefont{Itohi}} \bibnamefont{and}
  \bibinfo{author}{\bibfnamefont{M.}~\bibnamefont{Yamada}},
  \bibinfo{journal}{J. Electron. Mater.} \textbf{\bibinfo{volume}{38}},
  \bibinfo{pages}{925} (\bibinfo{year}{2009}).

\bibitem[{\citenamefont{Sadia and Gelbstein}(2012)}]{Gelbstein2012}
\bibinfo{author}{\bibfnamefont{Y.}~\bibnamefont{Sadia}} \bibnamefont{and}
  \bibinfo{author}{\bibfnamefont{Y.}~\bibnamefont{Gelbstein}},
  \bibinfo{journal}{J. Electron. Mater.} \textbf{\bibinfo{volume}{41}},
  \bibinfo{pages}{1504} (\bibinfo{year}{2012}).

\bibitem[{\citenamefont{Il'nitskaya et~al.}(1991)\citenamefont{Il'nitskaya,
  Bruskov, Zavalii, and Kuz'ma}}]{NiSi3P4Structure}
\bibinfo{author}{\bibfnamefont{O.~N.} \bibnamefont{Il'nitskaya}},
  \bibinfo{author}{\bibfnamefont{V.~A.} \bibnamefont{Bruskov}},
  \bibinfo{author}{\bibfnamefont{P.~Y.} \bibnamefont{Zavalii}},
  \bibnamefont{and} \bibinfo{author}{\bibfnamefont{Y.~B.}
  \bibnamefont{Kuz'ma}}, \bibinfo{journal}{Inorganic Materials (USSR)}
  \textbf{\bibinfo{volume}{27}}, \bibinfo{pages}{1108} (\bibinfo{year}{1991}).

\bibitem[{Cu3(1972)}]{Cu3SbSe4Structure}
\bibinfo{journal}{Acta Crystallogr. B} \textbf{\bibinfo{volume}{28}},
  \bibinfo{pages}{3672} (\bibinfo{year}{1972}).

\bibitem[{\citenamefont{Yang et~al.}(2011)\citenamefont{Yang, Huang, Wu, and
  Xu}}]{Cu3SbSe4Yang2011}
\bibinfo{author}{\bibfnamefont{C.}~\bibnamefont{Yang}},
  \bibinfo{author}{\bibfnamefont{F.}~\bibnamefont{Huang}},
  \bibinfo{author}{\bibfnamefont{L.}~\bibnamefont{Wu}}, \bibnamefont{and}
  \bibinfo{author}{\bibfnamefont{K.}~\bibnamefont{Xu}}, \bibinfo{journal}{J.
  Phys. D: Appl. Phys.} \textbf{\bibinfo{volume}{44}}, \bibinfo{pages}{295404}
  (\bibinfo{year}{2011}).

\bibitem[{\citenamefont{Skoug et~al.}(2011{\natexlab{a}})\citenamefont{Skoug,
  Cain, and Morelli}}]{Cu3SbSe4Skoug2011}
\bibinfo{author}{\bibfnamefont{E.~J.} \bibnamefont{Skoug}},
  \bibinfo{author}{\bibfnamefont{J.~D.} \bibnamefont{Cain}}, \bibnamefont{and}
  \bibinfo{author}{\bibfnamefont{D.~T.} \bibnamefont{Morelli}},
  \bibinfo{journal}{Appl. Phys. Lett.} \textbf{\bibinfo{volume}{98}},
  \bibinfo{pages}{261911} (\bibinfo{year}{2011}{\natexlab{a}}).

\bibitem[{\citenamefont{Skoug et~al.}(2011{\natexlab{b}})\citenamefont{Skoug,
  Cain, Majsztrik, Kirkham, Lara-Curzio, and Morelli}}]{Cu3SbSe4Skoug2011B}
\bibinfo{author}{\bibfnamefont{E.~J.} \bibnamefont{Skoug}},
  \bibinfo{author}{\bibfnamefont{J.~D.} \bibnamefont{Cain}},
  \bibinfo{author}{\bibfnamefont{P.}~\bibnamefont{Majsztrik}},
  \bibinfo{author}{\bibfnamefont{M.}~\bibnamefont{Kirkham}},
  \bibinfo{author}{\bibfnamefont{E.}~\bibnamefont{Lara-Curzio}},
  \bibnamefont{and} \bibinfo{author}{\bibfnamefont{D.~T.}
  \bibnamefont{Morelli}}, \bibinfo{journal}{Appl. Phys. Lett.}
  \textbf{\bibinfo{volume}{3}}, \bibinfo{pages}{602}
  (\bibinfo{year}{2011}{\natexlab{b}}).

\bibitem[{\citenamefont{Skoug et~al.}(2011{\natexlab{c}})\citenamefont{Skoug,
  Cain, Morelli, Kirkham, Majsztrik, and Lara-Curzio}}]{Cu3SbSe4Skoug2011C}
\bibinfo{author}{\bibfnamefont{E.~J.} \bibnamefont{Skoug}},
  \bibinfo{author}{\bibfnamefont{J.~D.} \bibnamefont{Cain}},
  \bibinfo{author}{\bibfnamefont{D.~T.} \bibnamefont{Morelli}},
  \bibinfo{author}{\bibfnamefont{M.}~\bibnamefont{Kirkham}},
  \bibinfo{author}{\bibfnamefont{P.}~\bibnamefont{Majsztrik}},
  \bibnamefont{and}
  \bibinfo{author}{\bibfnamefont{E.}~\bibnamefont{Lara-Curzio}},
  \bibinfo{journal}{J. Appl. Phys.} \textbf{\bibinfo{volume}{110}},
  \bibinfo{pages}{023501} (\bibinfo{year}{2011}{\natexlab{c}}).

\bibitem[{\citenamefont{Vick and Whittle}(1969)}]{BoronSilicon}
\bibinfo{author}{\bibfnamefont{G.~L.} \bibnamefont{Vick}} \bibnamefont{and}
  \bibinfo{author}{\bibfnamefont{K.~M.} \bibnamefont{Whittle}},
  \bibinfo{journal}{J. Electrochem. Soc.} \textbf{\bibinfo{volume}{116}},
  \bibinfo{pages}{1142} (\bibinfo{year}{1969}).

\bibitem[{\citenamefont{Goldsmid and Sharp}(1999)}]{EgEstimate}
\bibinfo{author}{\bibfnamefont{H.~J.} \bibnamefont{Goldsmid}} \bibnamefont{and}
  \bibinfo{author}{\bibfnamefont{J.~W.} \bibnamefont{Sharp}},
  \bibinfo{journal}{J. Electron. Mater.} \textbf{\bibinfo{volume}{28}},
  \bibinfo{pages}{869} (\bibinfo{year}{1999}).

\bibitem[{\citenamefont{Ioffe}(1957)}]{Ioffe}
\bibinfo{author}{\bibfnamefont{A.~F.} \bibnamefont{Ioffe}},
  \emph{\bibinfo{title}{Semiconductor Thermoelements and Thermoelectric
  Cooling}} (\bibinfo{publisher}{Infosearch Ltd}, \bibinfo{address}{London},
  \bibinfo{year}{1957}).

\bibitem[{\citenamefont{Toberer and Snyder}(2008)}]{ComplexTE}
\bibinfo{author}{\bibfnamefont{E.~S.} \bibnamefont{Toberer}} \bibnamefont{and}
  \bibinfo{author}{\bibfnamefont{G.~J.} \bibnamefont{Snyder}},
  \bibinfo{journal}{Nature Materials} \textbf{\bibinfo{volume}{7}},
  \bibinfo{pages}{105} (\bibinfo{year}{2008}).

\end{thebibliography}
\end{document}